\documentclass[twocolumn]{revtex4}
\usepackage{amssymb}
\usepackage{amsmath}
\usepackage{graphicx}
\usepackage{epsfig}
\usepackage[colorlinks]{hyperref}

\begin{document}

\title{\bf Magnetic and thermodynamic properties of Cu$_x$TiSe$_2$ single crystals}

\author{ Z.\,Pribulov\'{a},$^1$ Z.\,Medveck\'{a},$^1$ J.\,Ka\v{c}mar\v{c}\'{i}k,$^1$  V.\,Komanick\'{y},$^1$ T.\,Klein,$^2$, P.\,Rodi\`ere,$^2$ F.\,Levy-Bertrand,$^2$ B.\,Michon,$^2$ C.\,Marcenat,$^3$ P.\,Husan\'{i}kov\'{a},$^4$ V.\,Cambel,$^4$ J.\,\v{S}olt\'{y}s,$^4$ G.\,Karapetrov,$^5$ S.\,Borisenko,$^6$ D.\,Evtushinsky,$^7$ H.\,Berger,$^8$ and P.\,Samuely$^1$}

\affiliation{$^1$ Centre of Low Temperature Physics, Institute of Experimental Physics, Slovak Academy of Sciences, and P. J. \v{S}af\'{a}rik University, SK-04001 Ko\v sice, Slovakia\\
$^2$ Univ. Grenoble Alpes, Inst. NEEL, F-38042 Grenoble, France\\ 
$^3$ SPSMS, UMR-E9001, CEA-INAC/UJF-Grenoble 1, 17 Rue des martyrs, 38054 Grenoble, France\\
$^4$ Institute of Electrical Engineering, Slovak Academy of Sciences, D\'{u}bravsk\'{a} cesta 9, 84104 Bratislava, Slovakia\\
$^5$ Department of Physics, Drexel University, 3141 Chestnut St., Philadelphia, PA 19104, USA\\
$^6$ IFW Dresden, P.O. Box 270116, 01171 Dresden, Germany\\
$^7$ Helmholtz-Zentrum Berlin für Materialien und Energie, Albert-Einstein-Strasse 15, D-12489 Berlin, Germany\\
$^8$ Institute of Physics, Ecole Polytechnique F\' ed\' erale de Lausanne (EPFL), CH-1015 Lausanne, Switzerland}

\begin{abstract}
We present a detailed study of the phase diagram of copper intercalated TiSe$_2$ single crystals, combining local Hall-probe magnetometry, tunnel diode oscillator technique (TDO), and specific heat and angle-resolved photoemission spectroscopy measurements. A series of the Cu$_x$TiSe$_2$ samples from three different sources with various copper content $x$ and superconducting critical temperatures $T_c$ have been investigated. We first show that the vortex penetration mechanism is dominated by geometrical barriers enabling a precise determination of the lower critical field, $H_{c1}$. We then show that the temperature dependence of the superfluid density deduced from magnetic measurements (both $H_{c1}$ and TDO techniques) clearly suggests the existence of a small energy gap in the system, with a  coupling strength $2\Delta_s \sim [2.4-2.8]k_BT_c$, regardless of the copper content, in puzzling contradiction with specific heat measurements which can be well described by one single large gap $2\Delta_l \sim [3.7-3.9]k_BT_c$. Finally, our measurements reveal a non-trivial doping dependence of the condensation energy, which remains to be understood.
\end{abstract}
\maketitle

\section{Introduction}
The discovery of a charge ordered phase in underdoped cuprates \cite{CO-Cuprates} recently invigorated the debate on the origin of the coupling mechanism in high $T_c$ superconductors, which remains one of the major unsolved questions in solid state physics. As in many unconventional systems, the superconducting state develops in the vicinity of other electronic and/or magnetic instabilities and the interplay between superconductivity and those competing phases remains unclear. Dichalcogenides are then particularly interesting as they offer the opportunity to study this interplay in a much simpler system. Indeed, no competing magnetic instability develops in those systems but superconductivity still coexists with a charge density wave (CDW) instability. This interplay has first been studied into detail in 2$H$-NbSe$_2$ \cite{Valla} and, more recently, 1$T$-TiSe$_2$ became the focus of considerable interest as a (commensurate) CDW driven by an exciton-phonon mechanism \cite{Exciton} develops below $\sim 200$ K. This CDW is progressively suppressed upon Cu intercalation and recent x-ray diffraction measurements \cite{Xray-TiSe2} suggested that domain walls - associated with some (slight) incommensuration - appear for doping content over which a superconducting dome develops \cite{Morosan}. The influence of those domain walls remains to be understood, but the concomitant onset of superconductivity and incommensurability suggests that they may play a role in the formation of the superconducting state.

Moreover, despite its simple electronic structure \cite{elect_struc}, the nature of the superconducting gap(s) remains unclear in Cu$_x$TiSe$_2$. On one hand, thermal conductivity measurements \cite{Li 2007} suggested that this system is a conventional single-gap $s$-wave superconductor, in agreement with our recent specific-heat measurements \cite{Kacmarcik}. On the other hand, $\mu$SR measurements \cite{Zaberchik} displayed an anomalous temperature dependence of the London penetration depth indicating the presence of two superconducting gaps in underdoped Cu$_x$TiSe$_2$ where coexistence between CDW and superconductivity was anticipated. Recently, our local magnetic measurements revealed the existence of an unexpected transverse Meissner effect, clearly showing that vortices remain locked along the $ab-$planes in tilted magnetic fields \cite{lock-in}, hence indicating the presence of an unexpected - and still unexplained - strong modulation of the pinning energy along the $c-$direction, which might be related to a modulation of the gap/order parameter. 

In order to shed light on the nature of the superconducting properties, we performed a detailed study of the phase diagram of copper-intercalated TiSe$_2$ single crystals, combining local Hall-probe magnetometry (HPM), tunnel diode oscillator (TDO) technique and specific-heat measurements. We present a quantitative analysis of both the temperature and doping dependencies of the critical fields ($H_{c1}$ and $H_{c2}$), and hence of the corresponding penetration depth, $\lambda$ and coherence length, $\xi$ as well as the doping dependence of the superconducting gap(s). All the measurements demonstrate very good quality of the single crystals which all display well defined specific heat anomalies and very small pinning. We show that the vortex penetration mechanism is dominated by geometrical barriers which enables a reliable determination of $H_{c1}$. Those measurements, however, revealed a puzzling discrepancy between thermodynamic and magnetic properties. Indeed, whereas the former indicate the presence of one single large gap $2\Delta_l \sim [3.7-3.9]k_BT_c$, the temperature dependence of the superfluid density deduced from magnetic measurements (both HPM and TDO) is driven by a small gap $2\Delta_s \sim [2.4-2.8]k_BT_c$ at low temperatures. Finally, we show that the  condensation energy density calculated extracting $\lambda$ from $H_{c1}$ and $\xi$ from $H_{c2}$ measurements is consistent with previous measurements of the heat capacity; however, its temperature dependence is found to be nontrivial.

\section{Sample preparation and experiments}

Single crystals were prepared via the iodine gas transport method \cite{Oglesby}. Energy dispersive x-ray spectroscopy (EDS) analysis was performed to determine the copper content in the samples. The critical temperature $T_c$ of each sample, determined from the specific heat measurements, is displayed in the inset of Fig.1 together with the overall phase diagram previously suggested  by Morosan $et$ $al.$ \cite{Morosan}.  Samples A, B, C and D were grown in Karapetrov' s group, samples E and F by Berger and sample G by Levy-Bertrand and Michon. This latter sample is optimally doped, with the highest critical temperature, samples B, C and D are underdoped, while  samples A, E, and F are overdoped. The large collection of crystals hence made it possible to study the superconducting properties over a large part of the superconducting dome. The characteristic parameters deduced from our work have been summarized in Table I.

\begin{figure}[t]
\begin{center}
\resizebox{0.45\textwidth}{!}{\includegraphics{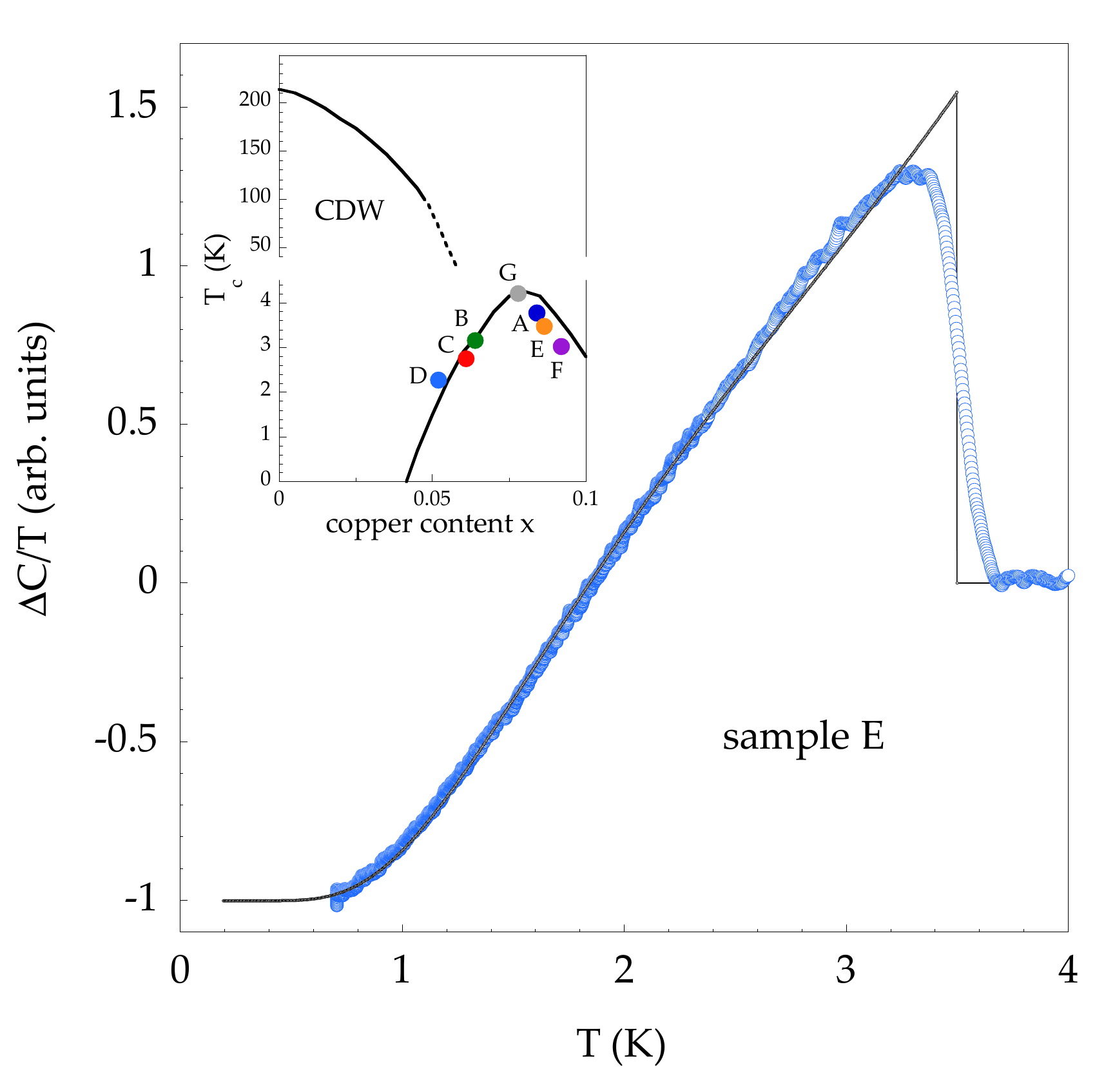}}
\caption{Temperature dependence of the electronic contribution to the specific heat of sample E. The solid line is a theoretical dependence for 2$\Delta$/k$T_c \sim$ 3.7. Inset: $T_c$ as a function of the copper content as proposed by Morosan {\it et al.} (Ref. \cite{Morosan}, solid lines) together with the critical temperatures of the samples studied in the present work (large symbols).}
\label{fig:fig2}
\end{center}
\end{figure}

 \begin{figure}[t]
\begin{center}
\resizebox{0.45\textwidth}{!}{\includegraphics{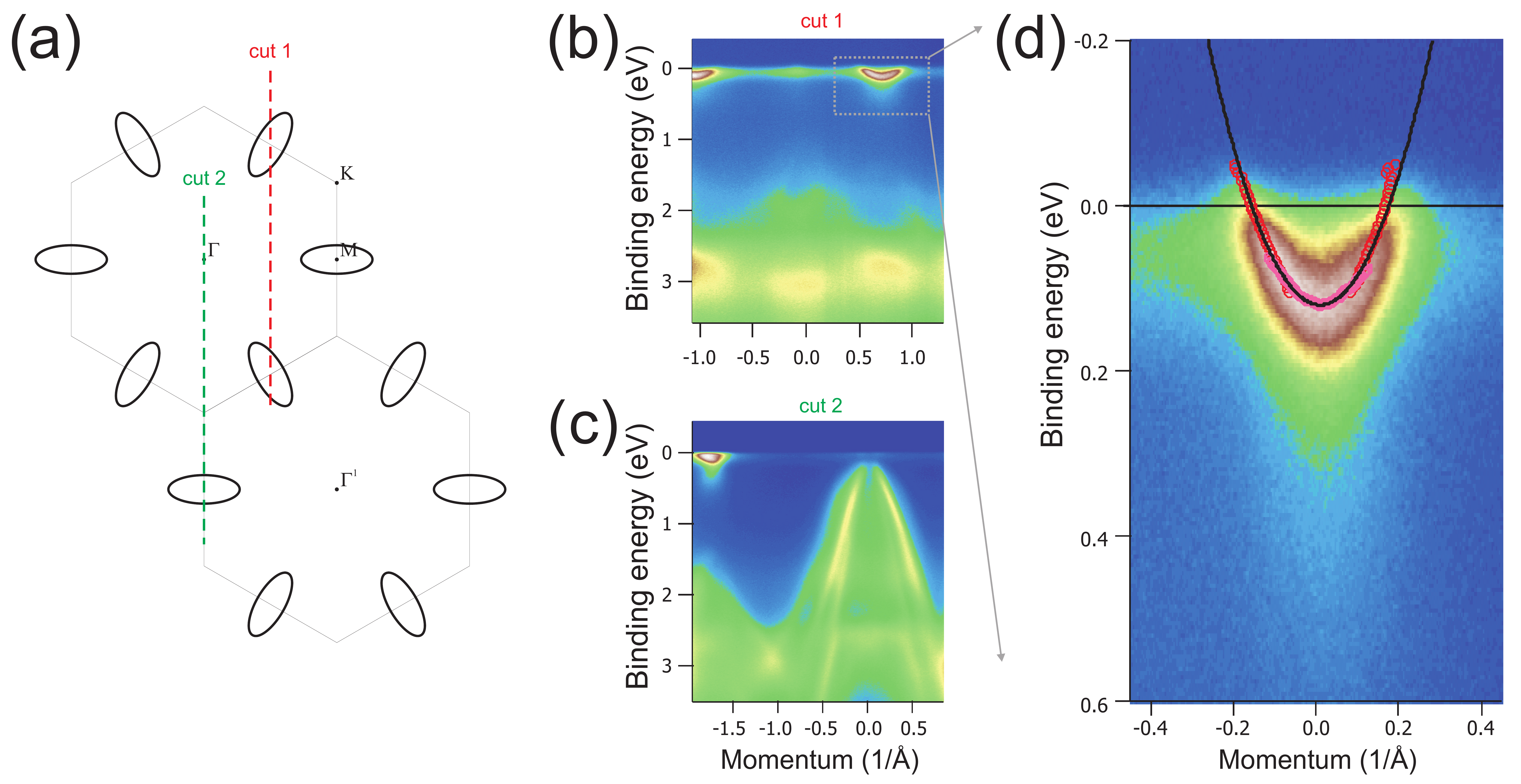}}
\caption{ARPES measurements of the electronic structure for Cu$_{0.07}$TiSe$_2$. (a) Schematic representation of the observed Fermi surface. (b,c) Energy-momentum cuts through the photoemission intensity distribution along the lines indicated in the panel (a). (d) Determination of the band dispersion in the vicinity of the Fermi level.}
\label{fig:fig2}
\end{center}
\end{figure}

\begin{table*}[t]
\caption{\label{table1}  Cu-doping content $x$, sample label, critical temperature $T_c$ (deduced from specific-heat measurements), first penetration field $H_p$ corresponding to the onset of the field penetration (for $T \rightarrow 0$, as deduced from local Hall probe measurements), $\alpha$ coefficient for geometrical barriers \cite{Brandt} and corresponding lower critical field $H_{c1}=\alpha H_p$, zero temperature upper critical field $H_{c2}$ (deduced from specific heat measurements \cite{Kacmarcik}), penetration depth $\lambda$, and coherence length $\xi$ deduced from the critical fields and $\kappa = \lambda/\xi$. Small gap values $\Delta_s$ (deduced from TDO measurements), and large gap values $\Delta_l$ (deduced from the temperature dependence of the specific heat).}
\begin{ruledtabular}
 \begin{tabular}{c c c c c c c c c c c c} 
 x & label & $T_c$(K) & $H_p$(0)(G) & $\alpha$ & $H_{c1}^c$(0)(G) & $H_{c2}^c$(0)(kG) & $\lambda_{ab}$(0)(nm) & $\xi_{ab}$(0)(nm) & $\kappa$(0) & $\Delta_s$(0)(K) & $\Delta_l$(0)(K)\\  [0.5ex] 
 \hline
0.052 & D & 2.3 & - & - & - & 5.0 & - & 25.5 & - & - & 4.3\\
\hline
 0.061 &  C & 2.8 & 14 & 4.3 & 60 & 5.5 & 290 & 24.5 & 11.8 & 4 & 5.2  \\ 
 \hline
 0.064 & B & 3.2 & 18  & 4.5 & 80 & 7.0 & 250 & 21.7 & 11.5 & 4 & 5.9 \\
 \hline
 0.075 & G & 4.1 & 50 & 2.3 & 115 & 9.5 & 207 & 18.6 & 11.1 & 4.9 & - \\
 \hline
 0.084 & A & 3.8 & 30 & 3.8 & 110 & 7.5 & 208 & 21 & 9.9 & 4.6 & 7 \\
 \hline
 0.086 & E & 3.5 & 21& 5.0 & 105 & 5.5 & 207 & 24.5 & 8.5 & - & 6.5 \\ 
\hline
 0.092 & F & 3.0 & 30 & 3.1 & 95 & 4.5 & 215 & 27 & 8.0 & - & 5.5 \\ [0.5ex] 
\end{tabular}
\end{ruledtabular}
\end{table*}

\begin{figure}[t]
\begin{center}
\resizebox{0.45\textwidth}{!}{\includegraphics{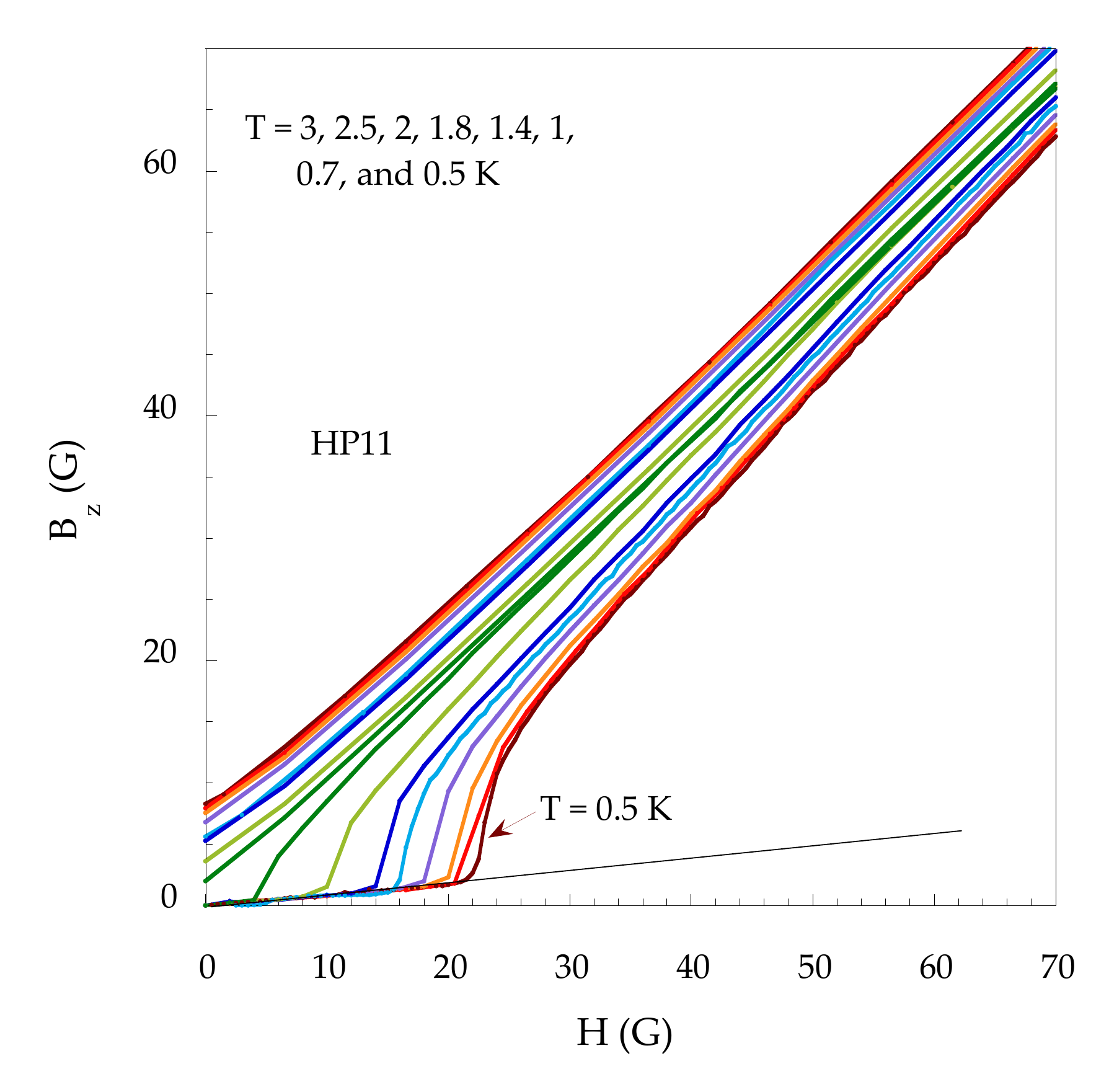}}
\caption{Magnetic field dependence of the local induction $B_z$ measured close to the center of sample E, at the indicated temperatures (different temperatures are marked by different colours). The field has been gradually increased up to 100 G and then decreased back to 0. The line marks the initial linear slope that is subtracted before further data treatment (see text). $H_p$ is unambiguously determined as the applied field above which $B_z$ departs from this linear behavior.}
\label{fig:fig3}
\end{center}
\end{figure}

Angle-resolved photoemission spectroscopy (ARPES) measurements were performed using "1-cubed" station at BESSY (Berliner Elektronenspeicherring-Gesellschaft für Synchrotronstrahlung) on a sample from the series grown by Berger. The doping level of the inspected sample, determined from the Fermi surface area is 0.07. From the measurements we infer the band dispersion and the Fermi surface shape (Fig.2). The Fermi surface consists of the approximately elliptic electron-like sheets centered around the M points. The observed ratio of the elliptical axes is about 2.5, and the depth of the band is about 120 meV. The measurements were performed with photon energy of 80 eV (Fig.2b) and 110 eV (Fig.2c). In both cases the sample temperature was 7 K. The orientation of the analyzer slit is given in Fig. 2(a).

Although the TiSe$_2$  is a layered compound, the Fermi surface, according to the bands structure calculations,  is substantially three-dimensional. In ARPES measurements the large degree of threedimensionality is confirmed by the smearing observed in the spectra. Both from theory and from experiment we estimate that the observed (maximal) depth of the band and size of the Fermi surface are effectively halved by the presence of the interlayer ($k_z$) dispersion. Uncertainty in this parameter is the main source of possible errors in the calculations based on the ARPES data.

The local magnetic field has been measured by placing the samples on top of high sensitivity ($\sim$ 1k$\Omega$/T) Hall-sensors patterned in epitaxial GaAs/AlGaAs heterostructures, forming 2D quantum wells. The magnetic field $H_a$ was applied perpendicularly to the sample basal planes ($ab$). The Hall probe arrays with 10x10 $\mu$m$^2$ active area and pitch ranging from 35 to 25 $\mu$m have been used to determine the field distribution over a length span of $\sim$ 300 $\mu m$. Depending on the sample dimensions, the crystal was shifted several times along the sensor line and a partial profile was recorded for every position. The complete profile has then been reconstructed by superimposing all partial measurements. Figure 3 displays, as an example, the magnetic field dependence of the local field, $B_z$, measured on a probe located close to the center of the sample (see discussion below) for the indicated temperatures, in a magnetic field perpendicular to the sample planes. 

 In the Meissner state, no magnetic field penetrates into the crystal but even minute distance between the sample and the probe gives rise to a small initial slope, as indicated in Fig.3. This contribution has been removed prior to any further data treatment. The number of vortices in the sensor area - and, correspondingly, the local magnetic field $B_z$ detected by the probe -  suddenly starts to grow when the applied field, $H_a$, reaches the first penetration field, $H_p$. Finally, some flux remains trapped in the sample when $H_a$ is turned back to zero leading to a finite remanent $B$ value. This remanent field indicates the presence of some bulk pinning. Taking $B \sim 5$ G over a sample width $\sim 100  \mu$m, one obtains a very small critical current on the order of 500 A/cm$^2$, highlighting the very good quality of the samples. The onset of the field penetration is very sharp and the presence of a small critical current does not put in question the determination of $ H_p$. Note that an anomalous transverse Meissner effect has been observed for the tilted magnetic fields in the samples C, G and A \cite{lock-in} (labeled sample 1, 2 and 3, respectively). 
 
 In the TDO measurements, the samples were attached to the end of a sapphire rod which was introduced in a coil of inductance $L$. Due to mutual inductance between the sample and the coil the resonant circuit of the LC oscillator (where L represents an inductor and C a capacitor) driven by the tunnel diode changes with the variation of the magnetic state of the sample. The variation of the magnetic penetration depth induces a change in $L$ and hence a shift of the resonant frequency $\delta f(T)=f(T)-f(T_{min})$ of a LC oscillating circuit (14 MHz) driven by a tunnel diode. This shift, renormalized to the one corresponding to the extraction of the sample from the coil ($\Delta f_0$) is then equal to the volume fraction ($\delta V/V$) of the sample which is penetrated by the field \cite{Diener}. For $H\|c$, $\delta V$ is related to the in-plane penetration depth $\lambda_{ab}$ through some calibration constant that depends on the sample geometry. However, this constant can be altered by edge roughness effects (see discussion in \cite{Klein}) introducing a wrong temperature dependence of the magnetic penetration depth. To avoid this, we have hence decided to perform all TDO measurements with $H\|ab$. In the following we show that even in this configuration the magnetic penetration depth probed is characteristic of $\lambda_{ab}$. Indeed, the surfaces parallel to the $ab$-planes are much flatter and $\delta V/V$ is, in this case, directly given by $\delta V/V \sim 2(\lambda_{c}/w+\lambda_{ab}/d) \sim 2/d\times[\lambda_{ab}+(d/w)\lambda_c]$ without any geometrical correction ($\lambda_c$ being the penetration depth parallel to the c-axis and $d$ and $w$ sample thickness and width respectively). Since $d/w<<1$, $\lambda_{ab}+(d/w)\lambda_c \sim \lambda_{ab}$ in this weakly anisotropic system \cite{ Kacmarcik}. A typical temperature dependence of the frequency shift in the TDO measurements up to $T_c$ is displayed in the inset of Fig.6 showing a very sharp decrease of $\Delta f$ for $T<T_c$, highlighting the high quality of the measured samples. 

Finally, specific-heat measurements have been performed  using an ac technique, as described elsewhere \cite{Sullivan,Kacmarcik}. The ac-calorimetry technique consists of applying a periodically modulated  power and measuring the resulting time-dependent temperature response. In our set-up, heating was provided by an optical fiber, and the temperature of the sample was recorded by a thermocouple, a precise $in situ$ calibration of the thermocouples in magnetic field was included in the data treatment. We performed measurements at temperatures down to 0.7 K and in magnetic fields up to 2 T. In this paper, only the measurements with the magnetic field oriented in the $c$ direction are considered. The electronic contribution to the specific heat $\Delta C_p/T=[C_p(T,H=0)-C_p(T,H>H_{c2})]/T$, together with the theoretical dependence for 2$\Delta/k_BT_c \sim$ 3.7 is displayed in Fig.1 in sample E, as an example. Very similar results were obtained in sample F (not shown here). The specific-heat anomaly at $T_c$ is very well resolved  in all samples, once again attesting for their high quality. For all samples $\Delta T_c/T_c$ is smaller than 0.08, $\Delta T_c$ being the transition width calculated between 10$\%$ and 90$\%$ of the specific heat anomaly. For the best sample, sample G, it is as small as 0.025.The specific-heat properties of samples A, B, C, and D were previously investigated into detail in \cite{Kacmarcik} (same sample labeling); specific heat of the sample G was presented in \cite{Levy}.

\section{Results and Discussion}
\subsection{Evidence for geometrical barriers in the vortex penetration mechanism}

\begin{figure}[t]
\begin{center}
\resizebox{0.45\textwidth}{!}{\includegraphics{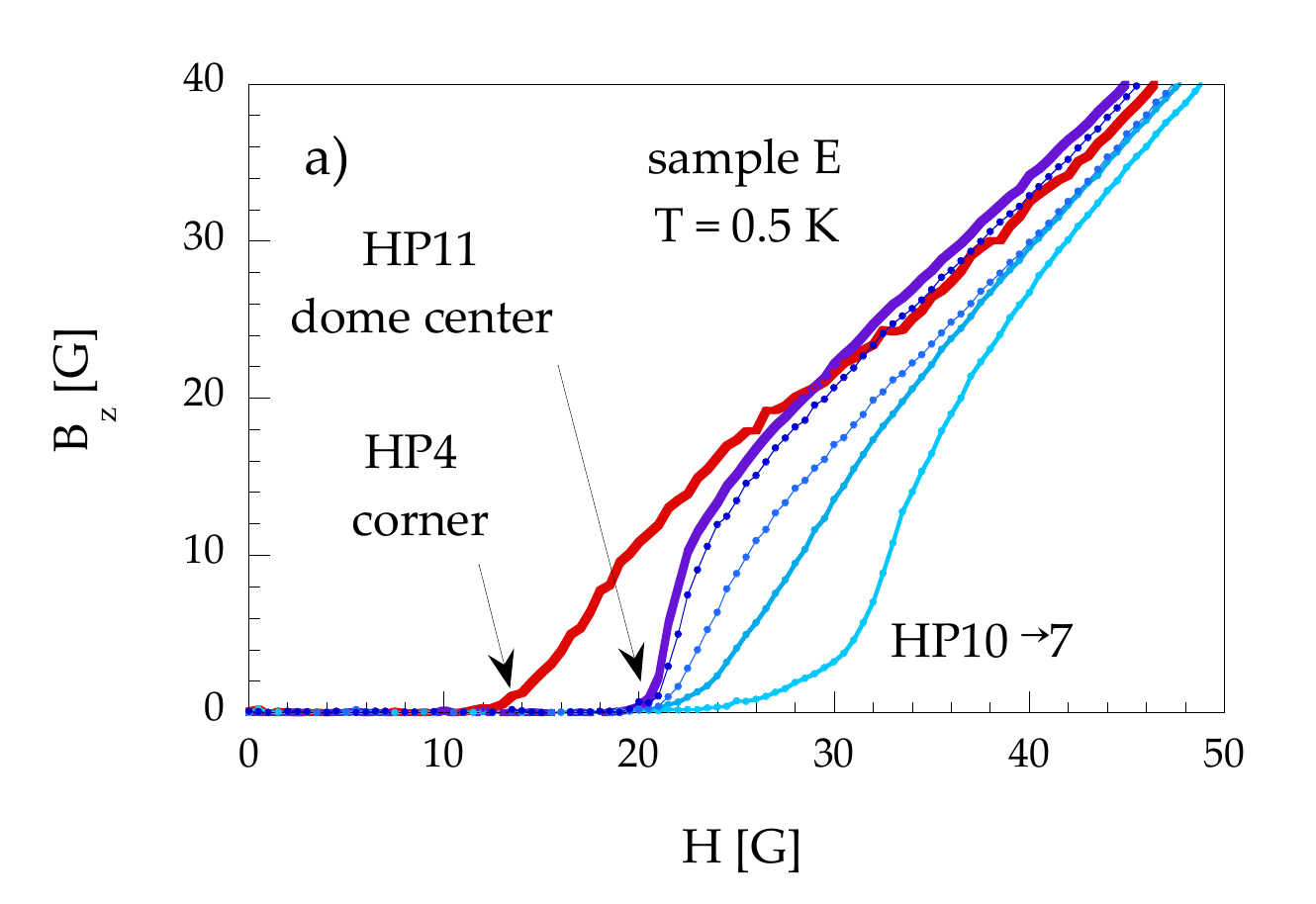}}
\resizebox{0.45\textwidth}{!}{\includegraphics{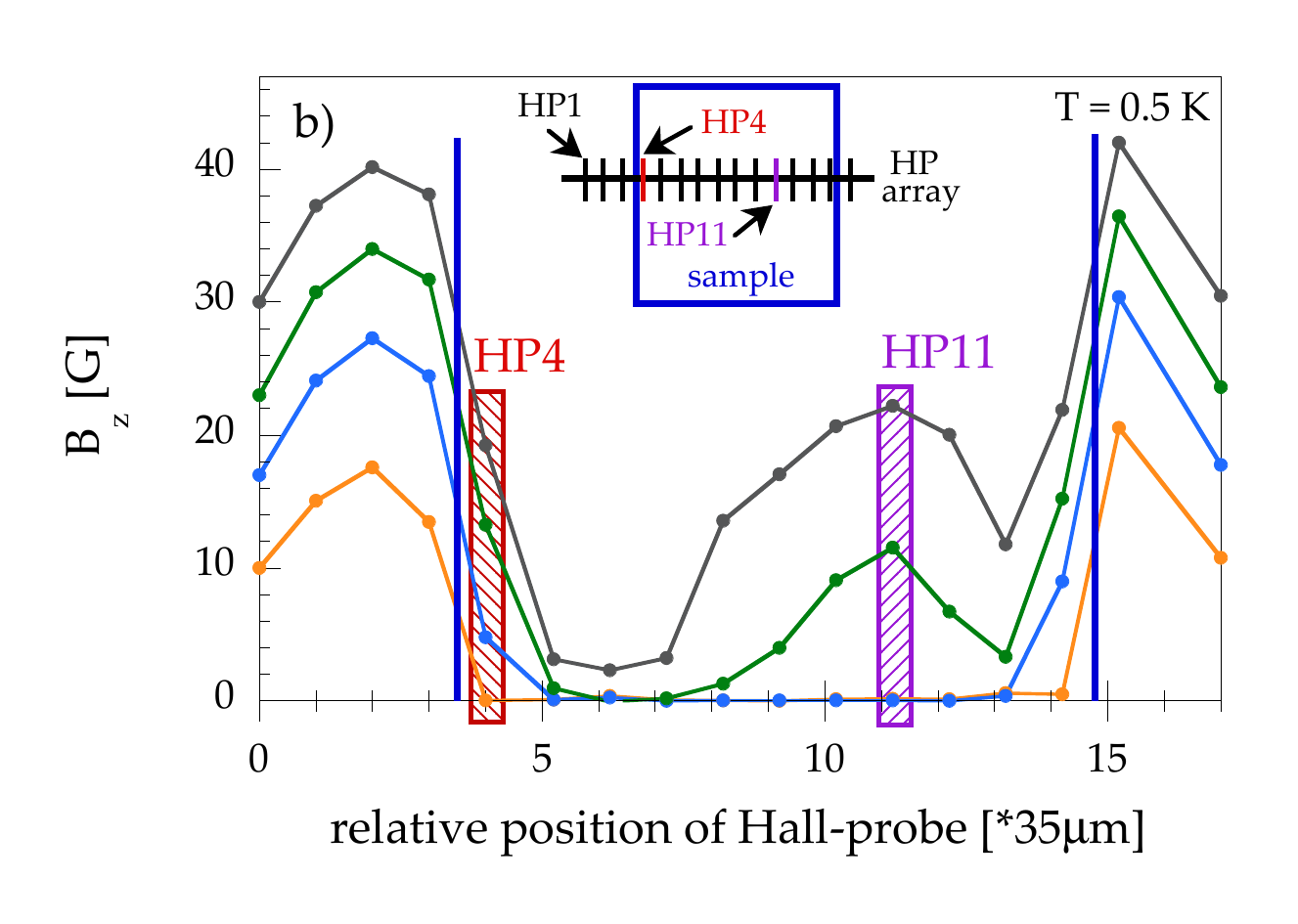}}
\caption{(a) $B$ as a function of applied magnetic field measurement at different probe positions, at $T=0.5$ K. The red thick line corresponds to Hall-probe No.4 located close to the sample edge and the purple thick line to Hall-probe Nos.11 corresponding to the center of the dome [see Fig.4b], dark blue to pale blue symbols/lines (from left to right) correspond to Hall-probes No.10, 9, 8, 7. Notation "dome center" and "corner" refers to Fig.4b, purple and red shaded boxes respectively. (b) Magnetic field profiles measured at 0.5 K in and around the sample for increasing applied magnetic field. Blue vertical lines indicate edges of the sample. Red and purple shaded boxes highlight the evolution of $B$ on HP4 and HP11 respectively [see red and purple lines in Fig4a]. The complete profile has been obtained by shifting the sample three times - see text for details. The right-most and left-most points in every profile correspond to the value of applied magnetic field.  Inset:  Sketch of the probe positions with respect to the sample.}
\label{fig:fig4}
\end{center}
\end{figure}

Figure 4a displays the induced magnetic field $B_z$ as a function of applied field $H_a$ in sample E (as an example) for several different Hall-probe positions [see sketch in the inset of Fig.4b].The spatial profile of induced magnetic field can be reconstructed by taking the $B_z$ values for each Hall-probe at a given $H_a$ value [main panel of Fig.4b]. For small $H_a$ values, the external field is shielded from all of the probes located below the sample (HP4 to HP14) and $B = 0$ (see the lowest -orange- profile for $H_a = 10~$G). As $H_a$ exceeds some critical value ($H_p\sim~20$ G), $B$ starts to increase more or less abruptly giving rise to a dome-like magnetic field profile (green and gray  profiles). This profile is characteristic of low pinning materials \cite{Zeldov} in the situation when the Meissner currents guide the vortices to the center of the sample. However, it is worth mentioning (see discussion below) that a partial penetration of the field is observed on HP4 and HP14 (edges on both sides of the sample) already for $H_a\sim$  13 G, i.e. for $H_a<<H_p$. Note also that the profile is slightly shifted towards the right side of the sample and the center of the dome does not match with the center of the sample. This is due to nonuniform distribution of the Meissner current density across the sample related to the slight thickness variation (the right side of the sample being slightly thinner).

In low pinning samples, the penetration process is determined by two main barriers: the Bean-Livingston barriers due to the attraction of penetrating vortices to the sample surface \cite{Bean-Livingston} and the geometrical barriers related to the nonelliptical shape of the plateletlike sample \cite{Zeldov,Brandt}. In the former case, $B=0$ in the whole sample for $H_a<H_p$ as the field penetrates only over a distance on the order of $\lambda$ (see discussion in Ref.\cite{Brandt}). On the contrary, in the presence of geometrical barriers, the magnetic field first penetrates partially through the sample corners, creating tilted vortices stuck in the edges. These partial field penetration regions expand from the corners both in  $z$ direction (perpendicular to the sample surface) and towards the sample center. Vortices finally jump to the center of the sample as the top and bottom parts meet at the equatorial point ($z=0$) for $H_a=H_p$. As shown in Fig.4a, this partial penetration in 
 the sample corners is clearly observed in our crystals, as a finite $B$ value is measured on probe HP4 (and HP14, not shown here) for $H_a$ values significantly smaller than for other probes, hence clearly indicating that the penetration process is dominated by geometrical barriers (see Refs.\cite{Pribulova} and \cite{lock-in} for a detailed analysis of the field dependence of the profiles in the framework of the geometrical barriers theory \cite{Zeldov}).

\begin{figure}[t]
\begin{center}
\resizebox{0.45\textwidth}{!}{\includegraphics{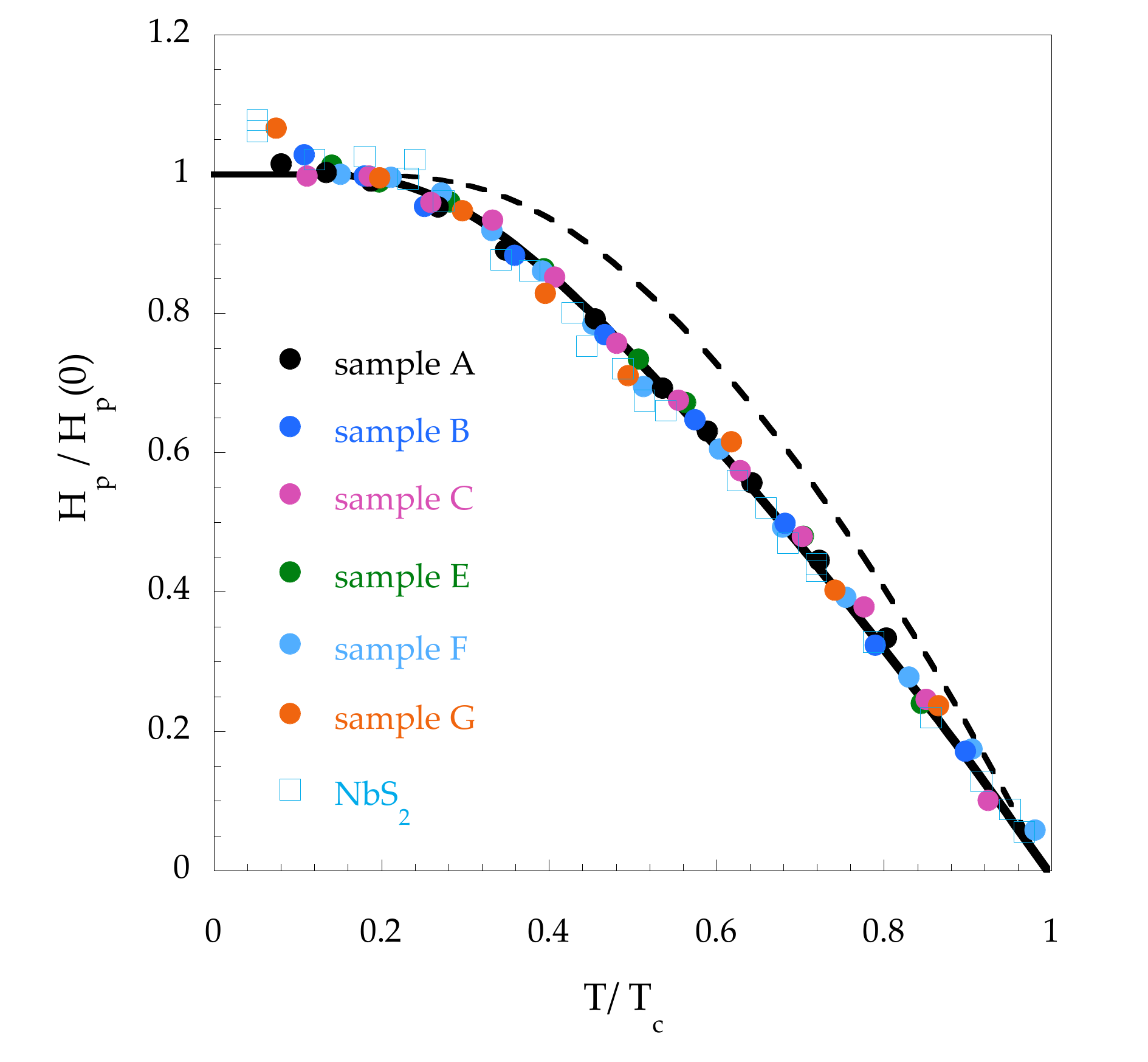}}
\caption{Temperature dependence of the first penetration field $H_p$ in several samples (in normalized scales). As shown, very similar temperature dependencies have been obtained in all measured samples. The dashed line is a theoretical dependence of $H_p$ corresponding to coupling ratio $2\Delta$~=~3.7~$k_BT_c$ and the thick line is the theoretical curve corresponding to presence of two energy gaps, 2$\Delta_l/k_BT_c\sim$3.7 and 2$\Delta_s/k_BT_c\sim$2.4, both with similar weight. Open symbols correspond to data previously obtained on NbS$_2$ (from \cite{NbS2}).}
\label{fig:fig5}
\end{center}
\end{figure}

\begin{figure}[t]
\begin{center}
\resizebox{0.45\textwidth}{!}{\includegraphics{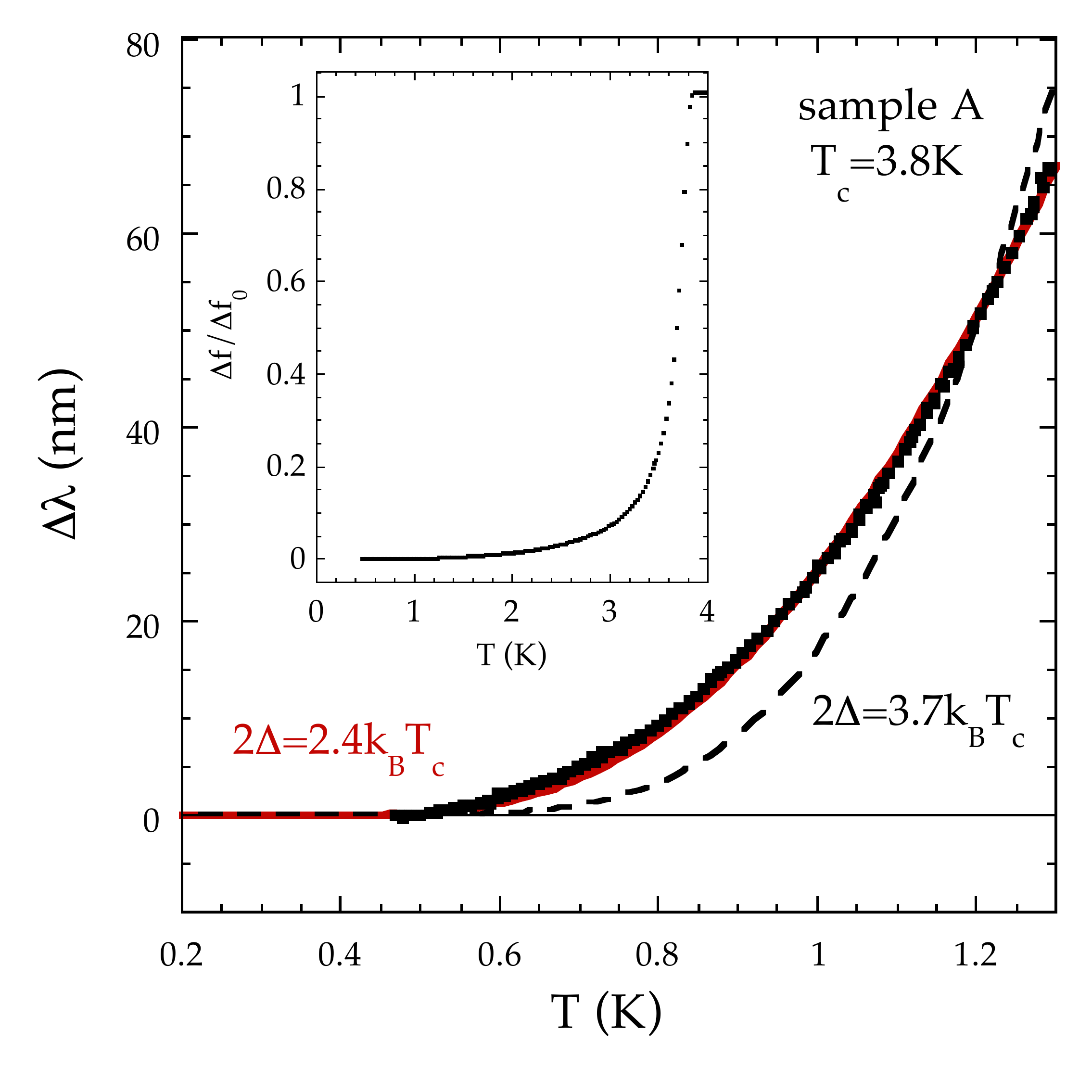}}
\caption{Temperature dependence of the penetration depth deduced from TDO measurements (sample A, as an example). As shown, $\lambda(T)$ can be well described by a standard exponential law with $2\Delta \sim 2.4k_BT_c$ (solid red line). On the other hand, an exponential law with $2\Delta \sim 3.7k_BT_c$ (as deduced from specific-heat measurements, see Fig.1)  leads to only very poor agreement with the experimental data (dashed black line). Inset: Temperature dependence of the frequency shift in the TDO measurements up to $T_c$ showing a sharp decrease of $\Delta f$ for $T<T_c$ highlighting the high quality of the measured samples.}
\label{fig:fig6}
\end{center}
\end{figure}

\subsection{Gap values}
In the presence of geometrical barriers, $H_p$ is directly proportional to $H_{c1}$, $H_p=\alpha H_{c1}$, where $\alpha$ is a geometrical factor depending on the sample thickness to width ratio \cite{Brandt2} and, neglecting a small temperature dependence of $\kappa = \lambda/\xi$, one  has:
$$\frac{H_p(T)}{H_p(0)}\approx\frac{\lambda^{2}(0)}{\lambda^{2}(T)}=1-2\int_{\Delta(T)}^{\infty} \frac{\partial f}{\partial E}\frac{E}{\sqrt{E^2-\Delta^2(T)}}dE$$
where $f $ is the Fermi function and $\Delta$ the superconducting gap. The corresponding temperature dependence of $H_p(T)/H_p(0)$ for all investigated samples is reported in Fig.5 (for normalized temperatures). As shown, the data can be well fitted introducing two energy gaps (thick line) in a simple $\alpha-$model \cite{Padamsee}:  2$\Delta_l/k_BT_c \sim$ 3.7 and 2$\Delta_s/k_BT_c \sim$ 2.4, both with similar weight. 

The presence of this small energy scale has been confirmed by TDO measurements. Indeed, as shown in Fig.6, the temperature dependence of the penetration depth can clearly be fitted by an exponential law attesting to the presence of a fully open superconducting gap with $2\Delta_s \sim 2.4k_BT_c$ (in sample A as an example). Very similar results have been obtained in sample B (2$\Delta_s/k_BT_c \sim2.5$), C (2$\Delta_s/k_BT_c \sim2.8$), and G (2$\Delta_s/k_BT_c \sim2.4$), attesting to the presence of a small gap for all doping contents, in good agreement with the temperature dependence of the lower critical field. The presence of this small gap is, to some extent, consistent with the $\mu$SR data \cite{Zaberchik}. However, in contrast with the present measurements, which do not show any significant change of the coupling ratios with doping ($2\Delta_s \sim [2.4-2.8]k_BT_c$), the $\mu$SR experiments suggested a clear increase of the coupling ratio of the small gap with  Cu content, leading to the merging of the two energy scales for optimal doping. 

Surprisingly, the observation of this small gap in magnetic measurements is in striking contrast with the result obtained in the specific-heat measurements. Indeed, in \cite{Kacmarcik} some of us showed that the temperature dependence of the specific heat of samples A, B, C, and D can be well described, introducing one single coupling ratio 2$\Delta = [3.7-3.9]k_BT_c$ for all copper concentrations, in agreement with the thermal conductivity measurements \cite{Li 2007}, which were suggesting that this system is a conventional single-gap superconductor. This fact is further supported by our present heat capacity measurements on samples E (see Fig.1) and F (not shown here). Indeed, even if the presence of the small gap ($\Delta \sim k_BT_c$, with the contribution weight of less than $10\%$) can not be fully excluded, the temperature dependence of the electronic  specific heat $\Delta C_p/T=[C_p(T,H=0)-C_p(T,H>H_{c2}]/T$ can be well described by a single gap model with 2$\Delta \sim 3.7k
 _BT_c$ (see solid line in Fig.1). Note that taking $2\Delta \sim 3.7k_BT_c$  leads to only very poor agreement with the experimental TDO data (see dashed black line in Fig.6) or with the temperature dependence of $H_p$ (see dashed line in Fig.5). While TDO is a surface sensitive method, Hall probe measurements are probing bulk properties; thus the difference between the surface and the bulk cannot explain our findings.

The explanation of such a discrepancy remains missing but it is worth noting that the magnetic measurements are probing the gap structure in the $ab-$plane, whereas the specific-heat measurements are averaging the gap structure over all $k-$directions and that the specific-heat measurements are mainly sensitive to heavy quasiparticles ($\gamma \propto m^*$), whereas the magnetic measurements are mainly probing the light quasiparticles ($1/\lambda^2 \propto 1/m^*$). This suggests a strong anisotropy of the effective mass over the Fermi surface and that the amplitude of the gap is strongly related to the effective mass. However, the temperature dependence of $H_{c1}$ (and hence the gap distribution) is very similar to the one previously  observed in 2$H$-dichalcogenides (NbSe$_2$ or NbS$_2$ \cite{NbS2}, open symbols in Fig.5) despite very different electronic structures (see \cite{elect_struc} and \cite{elect_struc_NbSe2}, respectively). Note also that, in those later systems, the multi-gap structure observed in magnetic measurements has been confirmed by both specific heat \cite{Kacmarcik2} and tunneling spectroscopy \cite{SpectroNbS2} measurements. The influence of the presence of a CDW leading to a strong $k-$dependence of the electron phonon coupling \cite{Leroux1} (and hence of the gap) first seemed also to be excluded since this CDW is present only in NbSe$_2$ (and not NbS$_2$), but recent measurements clearly showed a strong softening of the phonon modes in some directions even in NbS$_2$ \cite{Leroux}. 

The origin of the superconducting dome in Cu-TiSe$_2$ remains unclear. First-principles calculations emphasiszed the possible role of electron-electron correlations in the coupling constant of TiSe$_2$ (and MoS$_2$ flakes) \cite{Das-EE}. On the other hand, x-ray experiments performed on TiSe$_2$ single crystals \cite{Joe} showed that the end point of the CDW region occurs for pressures ($\sim 5$ GPa) being much larger than the pressures over which a superconducting dome is observed ($\sim 2-3.5$ GPa). Thus the superconducting dome is probably not directly related to a quantum critical point corresponding to the vanishing of the CDW phase. On the other hand, some incommensurability of CDW was observed in the superconducting region, suggesting that superconductivity could be related to the formation of CDW domain walls. This idea is further supplemented by the observation of CDW incommensurability also in Cu-doped samples by x-ray \cite{Xray-TiSe2}
and by optical measurements \cite{Lioi}. Note that the correlation length of the CDW (the size of the incommensurable domains) in the $c$ direction was reported to be on the order of 22 unit cells \cite{Xray-TiSe2}, which is strikingly similar to the superconducting coherence length. In high-$T_c$ superconductors the lock-in effect accommodates due to the interlayer distance being larger than, or at least on the order of the coherence length. A domain superstructure with exactly this scale could be the reason for this effect in Cu$_x$TiSe$_2$. Note also that observation of the lock-in effect \cite{lock-in} in this system points to a strong variation of the line tension (i.e., superfluid density) along the $c$ direction. Even if this scenario requires further consideration, strong modulations of the superconducting parameter might probably lead to different  "gap measurements" depending on the probe used to determine this gap.

After 30 years of intensive research, the superconducting-gap structure still remains a hot topic in cuprates. Recently, Bru\'er $et$ $al.$ has shown that in YBa$_2$Cu$_3$O$_{7-\delta}$ the tunneling spectrum gets parallel contributions from a two-dimensional band structure where beside a conventional BCS $d$-wave pairing gap an additional small "gap" is revealed coming from unpaired states \cite{Bruer}. A large body of experimental data suggesting a coexisting two-gap scenario, i.e., superconducting gap and pseudogap, over the whole superconducting dome in several classes of cuprates has been collected \cite{Hufner}. While the latter scenario seems to be excluded in Cu$_x$TiSe$_2$ the detailed study of different  spatially sensitive channels are to be addressed.

\begin{figure}[t]
\begin{center}
\resizebox{0.5\textwidth}{!}{\includegraphics{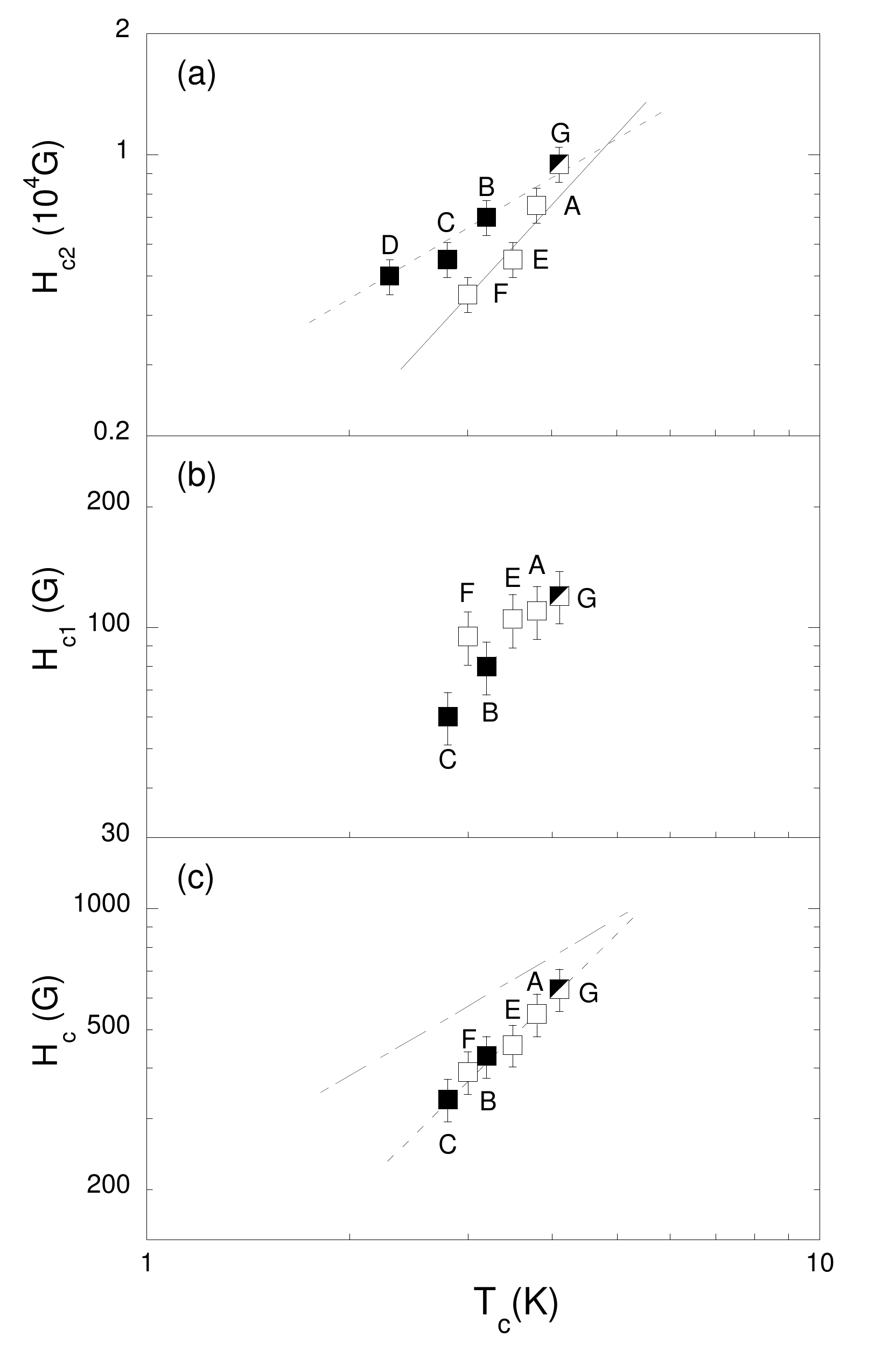}}
\caption{$T_c$ dependence of the upper [$H_{c2}$, panel (a)], lower [$H_{c1}$, panel (b)] and thermodynamic ($H_c$, panel (c)) critical fields in underdoped (closed symbols) and overdoped (open symbols) Cu$_x$TiSe$_2$ single crystals. The dotted and solid lines in panel (a) correspond to $T_c$ and $T_c^2$ dependencies, respectively, suggesting that underdoped samples are in the dirty limit whereas overdoped ones would be in the clean limit (see text for details). The dashed line in panel (c) indicates that   that $H_c$ scales as $T_c^{1.5}$ clearly deviating from the standard $T_c$ dependence (dot-dashed line). }
\label{fig:fig7}
\end{center}
\end{figure}

\subsection{Critical fields and condensation energy}

The critical fields $H_{c1}$ and $H_{c2}$, and corresponding values of penetration depth and coherence length of the different samples were collected and are listed in Table I. The $H_{c2}(0)$ values were derived from the specific-heat measurements. The data for crystals A, B, C, and D have been taken from \cite{Kacmarcik} and are supplemented by measurements on the other samples. The values of $H_{c1}(0)$ are directly derived from $H_p(0)$ introducing the $\alpha$ correction displayed in the table. 

In order to prove the accuracy of the $\lambda$ values deduced from our $H_p$ measurements we performed a thermodynamic consistency check (see also \cite{thermoconsistency}). The density of condensation energy, $\mu_0H_c^2/2$ is related to the density of states at the Fermi level $g(E_F)$ through $\mu_0H_c^2/2 \sim g(E_F)\Delta^2/2$ ($H_c = \Phi_0/(\mu_02\sqrt{2}\pi\lambda(0)\xi(0))$ being the thermodynamic field). Introducing the Sommerfeld coefficient ($\gamma$) through $\gamma=\pi^2k_B^2g(E_F)/3$ and taking $2\Delta \sim [3.7-3.9]k_BT_c$, one obtains $\gamma \sim (1.6\pm0.5).10^9/(\lambda(0)$[nm]$\xi(0)$[nm]$T_c$[K]$)^2 \sim 6\pm2$ mJ/molK$^2$ from our measurements. This value is in reasonable agreement with the measurements of the heat capacity performed by Morosan {\it et al.} \cite{Morosan}, giving $\gamma \sim 4-5$ mJ/molK$^2$, hence validating our results. 

Figure 7 displays an evolution of the critical fields with the critical temperature of the samples. As shown in panel (a), $H_{c2}$ roughly scales as $T_c^n$, with $n \sim 1$ (dashed line) and $\sim 2$ (solid line) for the underdoped  and the overdoped samples, respectively, suggesting that the system is in the dirty ($H_{c2}\propto 1/(\xi_0 l) \sim \Delta/(v_F l) \sim T_c$) and clean ($H_{c2}\propto 1/\xi^2 \sim \Delta^2/v_F^2 \sim T_c^2$) regimes, respectively. The $T_c$ dependence of $H_{c1}$ [panel (b)], and subsequently of $H_c$ (panel (c)), is much more puzzling. Indeed, in conventional superconductors, one expects that  $n\sim 1$ and $n \sim0$ for $H_{c1}$ in the dirty  and clean limits, respectively (in the dirty limit $\lambda \propto \lambda_L \sqrt{\xi_0/l}$ and $H_{c1}\propto /\lambda^2 \sim l/\lambda_L^2\xi_0 \sim  T_c$). Then, the $T_c$ dependencies for $H_{c2}$ and $H_{c1}$ lead to $H_c \propto \sqrt{H_{c1}H_{c2}} \sim T_c$ whatever the disorder. Our measurements, however, suggest that $H_c$ follows rather $H_c \propto T_c^{1.5}$ dependence; see dashed line in panel (c). For a comparison, the  $H_c \propto T_c$ is shown by the dash-dotted line as well. Such a surprising $T_c$ dependence of  $H_c$ has been reported in iron-based materials (see for instance \cite{Ni-doped}) and could be the signature of either a strong pair-breaking effect or the presence of superconducting quantum critical points in the vicinity of the end point of the superconducting dome. The former possibility can here be excluded as strong pair-breaking effects are expected to lead to some power-law dependence of the superfluid density, in striking contrast with our measurements.  Note that the important change in $H_{c1}$ with doping cannot be attributed to the change in the carrier concentration ($n \propto 1/\lambda^2$).

Finally, we compare values of $\lambda$, as well as $\xi$, derived from the lower and upper critical fields with those from ARPES measurements. ARPES supplies the London penetration depth $\lambda_L$ and the Pippard coherence length $\xi_0$ directly from electronic band structure. On the other hand, the quantities $\lambda$ and $\xi$ obtained from $H_{c1}$ and $H_{c2}$ are affected by the mean free path of the electrons, $l$. In the dirty limit they are related through formulas $\xi_0=\frac{\xi^2}{0.731l}$ and $\lambda_L=\lambda \sqrt{\frac{1.33l}{\xi_0}}$. 

In order to calculate $\lambda_L$ we need to take into account the shape of the Fermi surface  and Fermi velocity $v_F$  \cite{Evtushinsky}. Here we get an estimate of $\lambda_L=150\pm50$\,nm. Assuming the size of the superconducting gap $\Delta=0.6$\,meV, we estimate the coherence length $\xi_0$ using the formula $\xi_0 = \hbar v_F / (\pi \Delta) $, which results in $\xi_0=40\pm8$\,nm. These values were obtained on a sample with $x=0.07$, the concentration between those of samples B and G. In order to compare the samples with similar copper concentrations, we interpolated the values of the penetration depth and coherence length from Table I and obtained $\lambda=220$\,nm and $\xi=20$\,nm that would correspond to a sample with the same copper concentration as the one from ARPES.  Taking these values, using the formulas from above, we obtained consistent results of the mean free path $l=13.5$\,nm from both $\lambda$ and $\xi$. This confirms that such an underdoped sample is indeed in the dirty limit, as suggested by the evolution of $H_{c2}$. We would like to point out, that it is remarkable that in such a complicated system with CDW and shallow electronic bands, like Cu$_x$TiSe$_2$, we could arrive at very good agreement between two completely independent experimental approaches. It shows that a rather simple Fermi-liquid-like approach works well also in this complicated system, yet leaving an open question about where, in such smooth band structure, the two energy gaps could reside.

\section{Conclusions}

We have highlighted a surprising discrepancy between magnetic and thermodynamic measurements which led to seemingly contradictory results. Indeed, whereas the latter clearly suggested that this system is a conventional superconductor with $2\Delta_l \sim [3.7-3.9]k_BT_c$, the temperature dependence of the superfluid density (for $T\rightarrow 0$) deduced from magnetic measurements (both HPM and TDO) clearly shows the existence of a much smaller gap $2\Delta_s \sim [2.4-2.8]k_BT_c$. Our measurements are also pointing out a surprising dependence of the condensation energy density on $T_c$.

\acknowledgements

This  work  was supported  by EU ERDF (European regional development fund) Grant No. ITMS26220120005, by the Slovak Research and Development Agency,  under Grant No. APVV-14-0605, by the Slovak Scientific Grant Agency under Contract No. VEGA-0149/16 and VEGA-0409/15, by the U.S. Steel Ko\v sice, s.r.o., and by the French National ResearchAgency through Grant No. ANR-12-JS04-0003-01 SUBRISSYME.

\end{document}